\documentclass[%
 aip,
 jmp,%
 amsmath,amssymb,
 reprint,%
floatfix]{revtex4-1}

\usepackage{graphicx}
\usepackage{dcolumn}
\usepackage{bm}

\usepackage{color}

\begin{document}


\preprint{AIP/123-QED}

\title{Molecular dynamics study of the thermal conductivity in nanofluids}

\author{I. Topal}
\email{topali@itu.edu.tr}

\author{J. Servantie}
\email{cservantie@itu.edu.tr}

\affiliation{Department of Physics, Istanbul Technical University, Maslak 34469 \. Istanbul, Turkey}

\date{\today}

\begin{abstract}
We evaluate the thermal conductivity of a model nanofluid at various volume fractions of nanoparticles with equilibrium (EMD) 
and non-equilibrium (NEMD) molecular dynamics simulations. The Green-Kubo formalism is used for the EMD simulations while a net heat 
flux is imposed on the system for the NEMD simulations. The nanoparticle-nanoparticle, fluid-fluid and fluid-nanoparticle interactions
are all taken as Lennard-Jones potentials. An empirical parameter is added to the attractive part of the potential to control the
hydrophilicity of the nanoparticles, hence controlling how well dispersed are the nanoparticles in the base fluid. The results show that
the aggregation of the nanoparticles does not have a measurable effect on the conductivity of the nanofluid. Nanofluids with volume 
fractions of $2\%$ and $3\%$ show an enhanced conductivity with respect to the bulk fluid. Surprisingly, nanofluids with higher volume 
fractions did not show any enhancement of the conductivity.

\end{abstract}

\maketitle

\section{Introduction}

Nanofluids are defined as a base fluid containing well dispersed nano-sized solid particles. \cite{R1} Recent
experiments have suggested that nanofluids tend to have higher thermal conductivity than the base bulk fluids. \cite{R1-1}
There are few numerical studies of the thermal conductivity of nanofluids in the literature, one of the most prominent work
was performed by Sarkar \emph{et al.} \cite{9} They modelled a copper nanoparticle in
liquid argon using equilibrium molecular dynamics simulations (EMD). The Lennard-Jones potential was used to model both the fluid
and the nanoparticle. They evaluated the thermal conductivity of the nanofluid for a single copper
nanoparticle and varying volume fraction. The results suggest that the increase in thermal conductivity is mostly due to the
increased mobility of fluid atoms.\par

Sankar \emph{et al.} studied water-platinum nanoparticles nanofluid with EMD. \cite{S1} They used four different
interactions to have a more realistic nanofluid. They observed that the thermal conductivity of the nanofluid increases
proportionally with the temperature and volume fraction of the nanoparticle. Ghosh \emph{et al.} calculated the thermal conductivity
of water-copper nanofluids using a hybrid MD-stochastic model, \cite{10} they also observed a linear increase with the volume
fraction. Additionally,  Mohebbi \emph{et al.} \cite{P1} and Cui \emph{et al.} \cite{S3} also reported an increase in thermal
conductivity of nanofluids with the volume fraction of nanoparticles. On the other hand, some studies observed that the rate
of enhancement decreases with the volume fractions of nanoparticles, leading in some cases to a plateau
at a relatively small volume fractions of 2\% to 5\%. \cite{9,P2,P3}

{Cui \emph{et al.} observed that the thermal conductivity of nanofluids decreases as the nanoparticle diameter
increases.  \cite{rev2,rev1} However, depending on the type of nanoparticles increasing size can also lead to 
increasing thermal conductivity. \cite{R3} Another factor influencing the thermal conductivity is the shape of the nanoparticles. \cite{R1} Indeed, it was observed that 
higher surface to volume ratio of nanoparticles leads to a larger enhancement of the thermal conductivity. \cite{rev2}  Cui \emph{et al.} 
suggested that the shape of the nanoparticles has an impact on the radial distribution
function leading in turn to changes on the thermophysical properties of the nanofluid. \cite{rev5}}

Nanoparticle clustering is one of the mechanisms proposed for the enhancement of thermal conductivity. \cite{36}
Kang \emph{et al.}  studied nanoparticle aggregation with two nanoparticles and observed
that the thermal conductivity increases when the nanoparticles are close together. \cite{37} Similarly, Lee \emph{et al.} 
observed the aggregation of nanoparticles results in a higher increase of the thermal conductivity compared to well
dispersed nanoparticles. \cite{35} On the other hand, Sedighi \emph{et al.}  studied the thermal conductivity of a
water-silicon dioxide nanofluid and observed well dispersed nanofluids had a slightly larger enhancement of the thermal conductivity 
with respect to aggregated nanoparticles. \cite{P3}

Xue \emph{et al.}  studied the effect of layering on the thermal conductivity for a simple liquid
with non-equilibrium molecular dynamics simulations (NEMD). \cite{30} They did not observed any difference between the thermal conductivity of the
layered liquid and the bulk liquid and suggested to rule out layering as a mechanism for the enhancement of thermal conductivity in
nanofluids. Keblinski \emph{et al.} suggested that Maxwell's theory of well dispersed particles should be given up and allowed
chain-forming morphologies for nanoparticles so that the disagreement between the experiment and the effective medium theory
could be clarified. \cite{31} They mentioned the importance of aggregation on the thermal transport enhancement.

Babaei \emph{et al.} calculated thermal conductivity of different multi-component systems via the Green-Kubo formula
using EMD by comparing the results with the NEMD calculated results \cite{33}. They did not observe any significant
enhancement for well-dispersed nanofluid. They underlined the importance in correctly defining the average energies used in
the evaluation of the heat current. 

In this study, we use a generic coarse-grained model for the nanoparticles and a Lennard-Jones fluid for the base
fluid. The interactions strengths are varied in order to evaluate the effect of layering and aggregation on the thermal
conductivity. The volume fraction is also varied. The thermal conductivity is first evaluated with EMD then validated
with NEMD simulations.

The paper is organized as follows: In Sec. II, we outline the details of model we use in the study. Then, in
Sec. III, we compute the thermal conductivity for varying aggregations and  volume fractions of nanoparticles.
Finally conclusions are drawn in Sec. IV.

\section{\label{sec2} Model}
We are interested in studying the universal properties of a nanofluid, consequently we use a coarse
grained model. The base fluid is modeled as a Lennard-Jones fluid, hence, the fluid-fluid interactions
are described by a 6-12 Lennard-Jones potential, \cite{B1}
\begin{equation}
  \mathcal{V}_{\rm LJ}(r)=\left\{
  \begin{array}{cc}
    4\epsilon \left[\left(\frac{\sigma}{r}\right)^{12}-\left(\frac{\sigma}{r}\right)^6 \right] & \text{for $r < r_c$} \\
    0 & \text{ for $r > r_c$}
  \end{array}
  \right.
  \label{eqLJ}
\end{equation}
where the cut-off distance is taken as $r_c=2.5 \sigma$. We note that the same value is used for all the interactions 
in the model.
The nanoparticles are modelled as roughly spherical molecules with a radius of $r=2\sigma$. They consist of 58
atoms. The atoms inside nanoparticles interact with the Lennard-Jones potential of Eq. \ref{eqLJ}, and additionally,
with the Finitely Extensible Non-Linear Elastic (FENE)
potential, \cite{34}
\begin{equation}
  \mathcal{V}_{\rm FENE}(r)=
  \left\{
  \begin{array}{cc}
    \frac{1}{2} k R_0^2 \ln\left[1-\left(\frac{r}{R_0}\right)^2\right] & \text{ for $r < R_0$} \\
    \infty & \text{ for $r \geq R_0$}
  \end{array}
  \right.
\end{equation}
where $ R_0=1.5\sigma, k=30 \epsilon/\sigma^2 $. The nanoparticles are first constructed and equilibrated in a separate
molecular dynamics simulation, and afterwards are added to the bulk fluid. The bulk fluid and nanoparticles are mixed to
obtain $4$ simulation boxes with varying nanoparticle volume fractions of $2\%$, $3\%$, $6\%$ and $10\%$. We define the volume
fraction of the nanoparticles $\varphi$ as,
\begin{equation}
 \varphi = \frac{\frac{4}{3} \pi r_{p}^3 }{V}
\end{equation}
where $r_{p}$ is the radius of the nanoparticle and $V$ is the volume of the simulation box.
The nanoparticles interact with the fluid and with other nanoparticles through a modified Lennard-Jones
potential,
\begin{equation}
  \mathcal{V}_{\alpha \beta}(r)= \left\{
  \begin{array}{cc}
     4\epsilon \left[\left(\frac{\sigma}{r}\right)^{12}- \zeta_{\alpha \beta} \left(\frac{\sigma}{r}\right)^6 \right] & \text{ for $r < r_c$} \\
                     0 & \text{ for $r > r_c$}
  \end{array}
  \right. 
\end{equation}
where $\alpha,\beta=n,f$ denotes the interaction occurs between a nanoparticle atom ($n$) or a fluid atom ($f$). The
coefficient $\zeta_{\alpha \beta}$ controls the magnitude of the attractive part of the interaction,
large $\zeta_{\alpha \beta}$ corresponds to a hydrophilic interaction. In order to have a nanofluid we have to prevent
the nanoparticles from flocculating, and thus have a well dispersed fluid. Consequently, the interaction between
nanoparticles should be hydrophobic. We found that $\zeta_{nn}=0.3$ ensures the nanoparticles are well dispersed in
the fluid, and fixed its value in all simulations. On the other hand, $\zeta_{nf}$ permits to investigate how the thermal
conductivity is influenced by the hydrophilicity of the nanoparticles, and as a consequence of the density of the base
fluid in their vicinity. For this purpose, we use three different values of the hydrophilicity parameter of the
fluid-nanoparticle interaction, namely $\zeta_{nf}=0.5$, $\zeta_{nf}=1$ and $\zeta_{nf}=1.5$.

Initially, a total number of $5000$ fluid atoms are arranged in a regular FCC lattice. The system size in the $x$ and $y$
directions is $15\sigma$ and varies between $29-31.5\sigma$ in the $z$ direction in order to reach the same fluid density.
The equations of motion are then integrated with the Velocity Verlet algorithm with a time step of $\Delta t=0.001 \tau$.
The molecular dynamic code and the post-processing codes are all written in FORTRAN 90. The code is 
parallelized with the openMP protocol and each simulation was performed with 4 processors.
The total energy $E$ and the total momentum $P$ are computed to check the validity of the code since they are both conserved
quantities. Using the Lennard-Jones phase diagram from literature \cite{29}, we choose a phase point corresponding to
a liquid state. The temperature and density are taken respectively as $k_B T=1.1 \epsilon$ and $\rho=0.7798 \sigma^{-3}$.

\begin{figure}[h!]
    \centering
    \includegraphics[width=3cm]{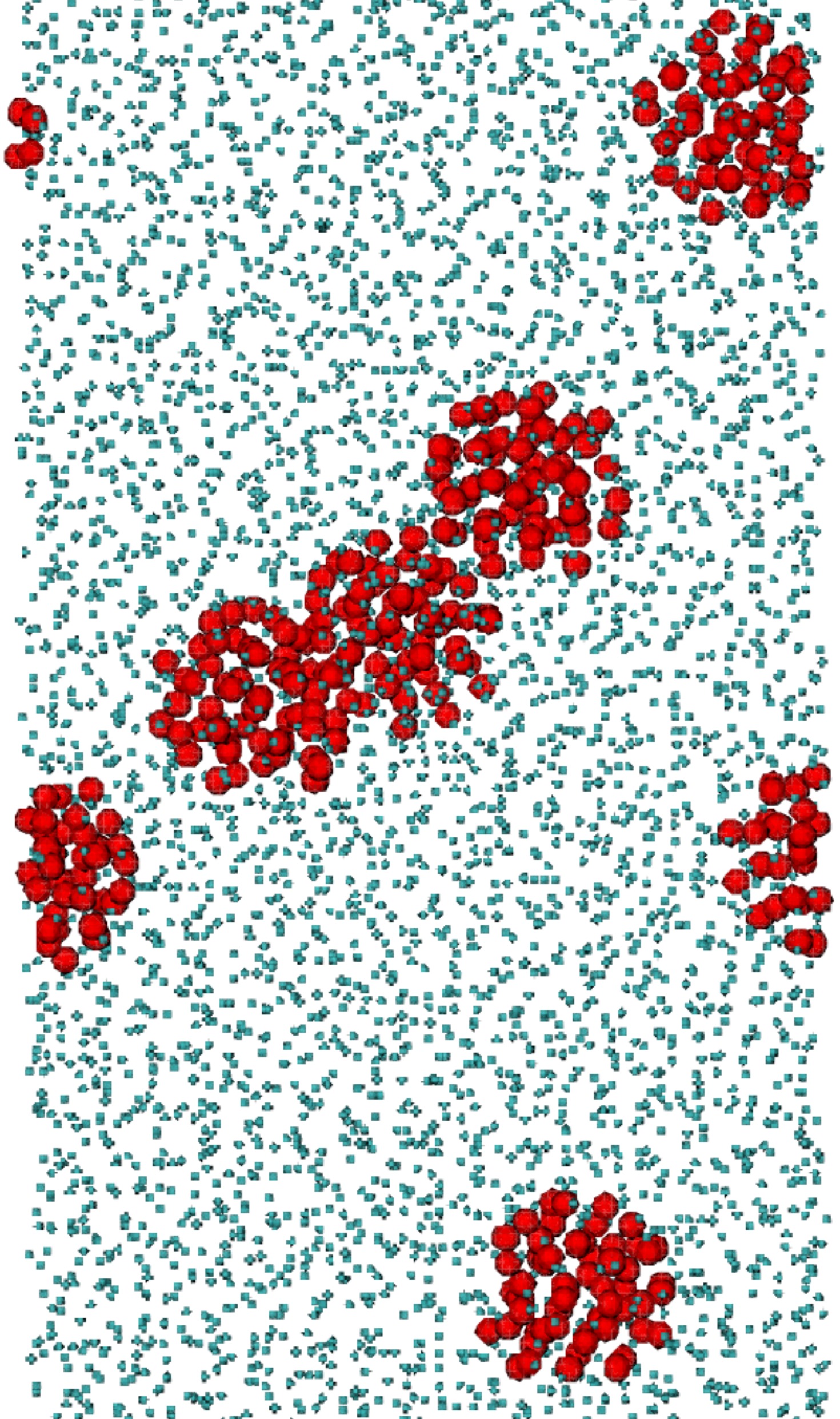}
    \includegraphics[width=3cm]{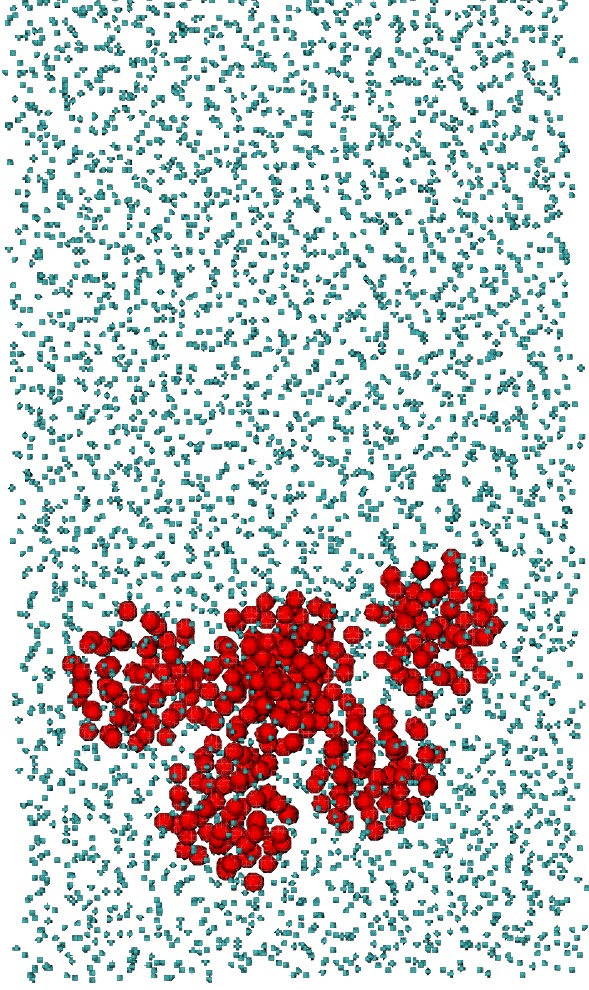}
    \caption{(Colour online) Two equilibrated nanofluids with nanoparticle volume fraction of $\%10$, on the left-hand-side $\zeta_{nf}$ = 1.5 
    and on the right-hand-side $\zeta_{nf}$ = 0.5.}
    \label{156np}
\end{figure}

For four different nanofluid models and three different Lennard-Jones potential, in total 12 sets, we initially run $10^7$
time steps to equilibrate the system, indeed it is known that in order to evaluate the transport properties of 
nanofluids a stable dispersion should be achieved. \cite{rev6} The resulting coordinates are then used for all the simulations. We depict in Fig. \ref{156np}
snapshots of nanofluids consisting of $6$ nanoparticles. The left-hand-side corresponds to $\zeta_{nf}=1.5$, a well dispersed nanofluid
whilst the right-hand-side has $\zeta_{nf}=0.5$ which yields an aggregated nanofluid.

\begin{figure}[t!]
 \centering
 \includegraphics[scale=0.35]{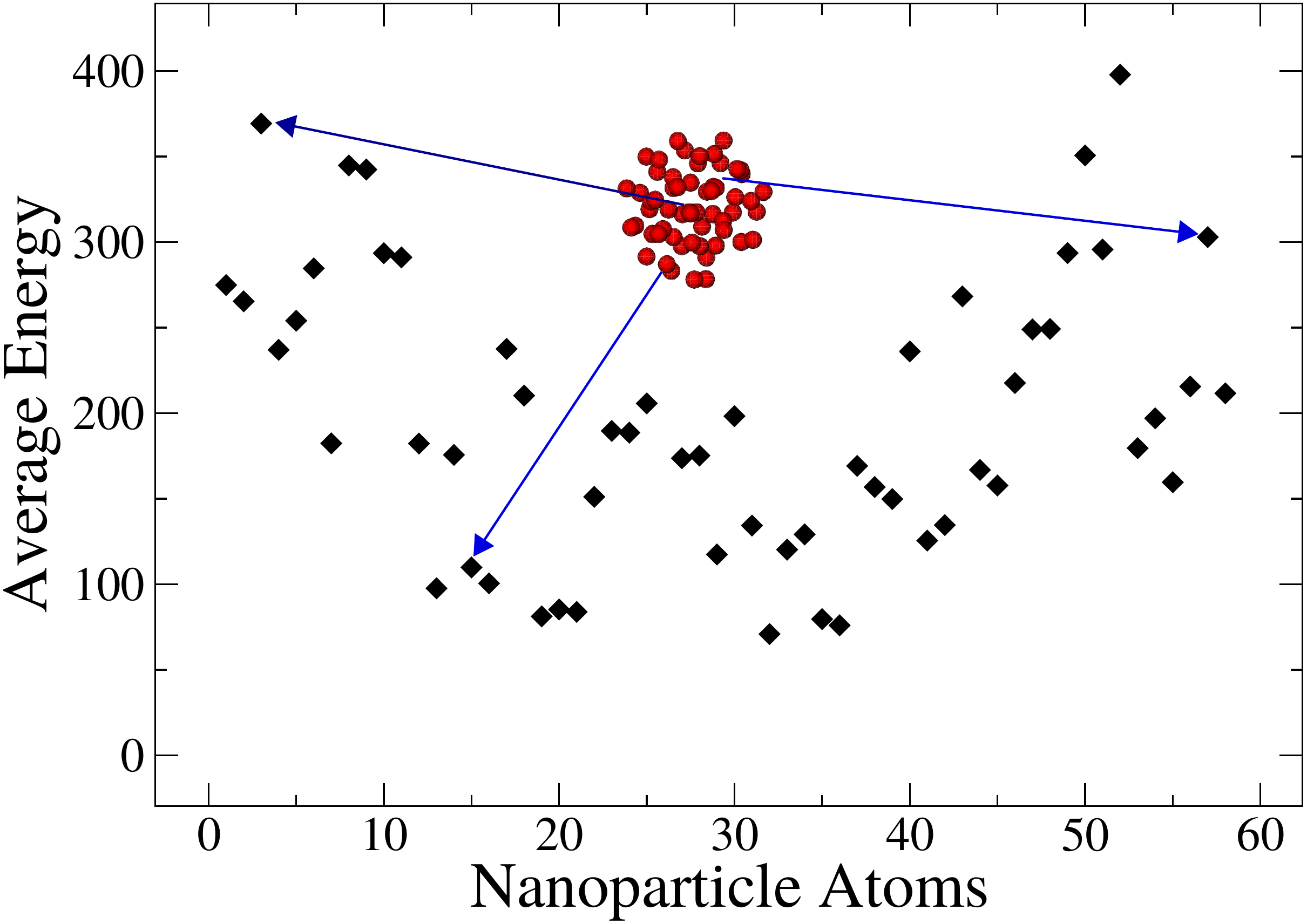}
 \caption{(Colour online)Average energies per atom inside a nanoparticle}
  \label{average}
\end{figure}

In order to check the consistency of our results we compute the thermal conductivity coefficient with two
different methods. First with equilibrium molecular dynamic simulation thanks to the Green-Kubo relation, \cite{39}
\begin{equation}
 \lambda = \frac{V}{3k_{B}T^2} \int_{0}^{\infty} d\tau \langle \mathbf{j}_{\lambda}(\tau) \mathbf{j}_{\lambda}(0)\rangle,
 \label{eq:gk}
\end{equation}
where $T$ is the temperature of the system, $V$ is the volume, $k_{B}$ the Boltzmann constant and $\mathbf{j}_{\lambda}$ the
microscopic heat current which is given by, \cite{36}
\begin{equation}
 \mathbf{j}_{\lambda} = \frac{1}{V} \left[\sum_{i=1}^{N} \mathbf{v}_{i}(E_{i} - \langle E_{i}\rangle) + \frac{1}{2} \sum_{i<j}^{N} \mathbf{r}_{ij}[\mathbf{F}_{ij}.(\mathbf{v}_{i} + \mathbf{v}_{j})]\right]
 \label{j}
\end{equation}
where $E_i$ is the instantaneous energy of the $i^{\rm th}$ atom,
\begin{equation}
 E_{i} = \frac{1}{2} m_{i} \mathbf{v}_{i}^{2} + \frac{1}{2} \sum_{j\neq i}^{N} \mathcal{V} (r_{ij}).
\end{equation}
One must take care in the calculation of the average individual energies $\langle E_i \rangle$. \cite{33} Indeed,
while for the bulk fluid one can simply take it to be the energy per particle, this does not hold for multi-component
systems anymore.  For the nanoparticles, as seen in Fig. \ref{average}, atoms are not identical anymore and the average energies per atom varies
wildly depending on the position of the atom inside the nanoparticle. Indeed, atoms in the nanoparticle which have
many neighboring atoms have larger energies with respect to atoms which have less neighboring atoms because of the
bonded potential. Consequently, before calculating the auto-correlation function in Eq. (\ref{eq:gk}), one must first
calculate the average energies. This is carried out by a preliminary simulation of $10^6$ time steps. Afterwards, the
heat current data is computed for a further $10^7$ time steps in order to compute the thermal conductivity.

The second method we use is a non-equilibrium molecular dynamics simulation. A gradient of temperature is imposed on the
system by applying stochastic boundary conditions in the $\hat{z}$ direction. Atoms going out of the simulation box
are reflected back with a random velocity corresponding to a Maxwell-Bolztmann distribution of a given temperature.
\cite{38} After a long  enough equilibration time a net heat current $\mathbf{j}_\lambda$ in the
$\hat{z}$ direction is established. Remark that because of surface effects the temperature profile has a non linear d
dependence, consequently the difference of temperature at walls does not correspond to the gradient of temperature. The
temperature profile is obtained by averaging the kinetic energies in slices of thickness $0.05\sigma$ as a function
of $z$. Linear fits to the data gives the gradients of temperatures, $\nabla T$, as a function of $\Delta T$. The imposed
heat current is calculated in the same way as for the Green-Kubo method and the thermal conductivity coefficient $\lambda$
is calculated using Fourier's law,
\begin{equation}
 \mathbf{j}_{\lambda} = -\lambda \nabla T
 \label{fl}
\end{equation}

\section{\label{sec3}Results}

\begin{figure}[h!]
 \centering
 \includegraphics[scale=0.3]{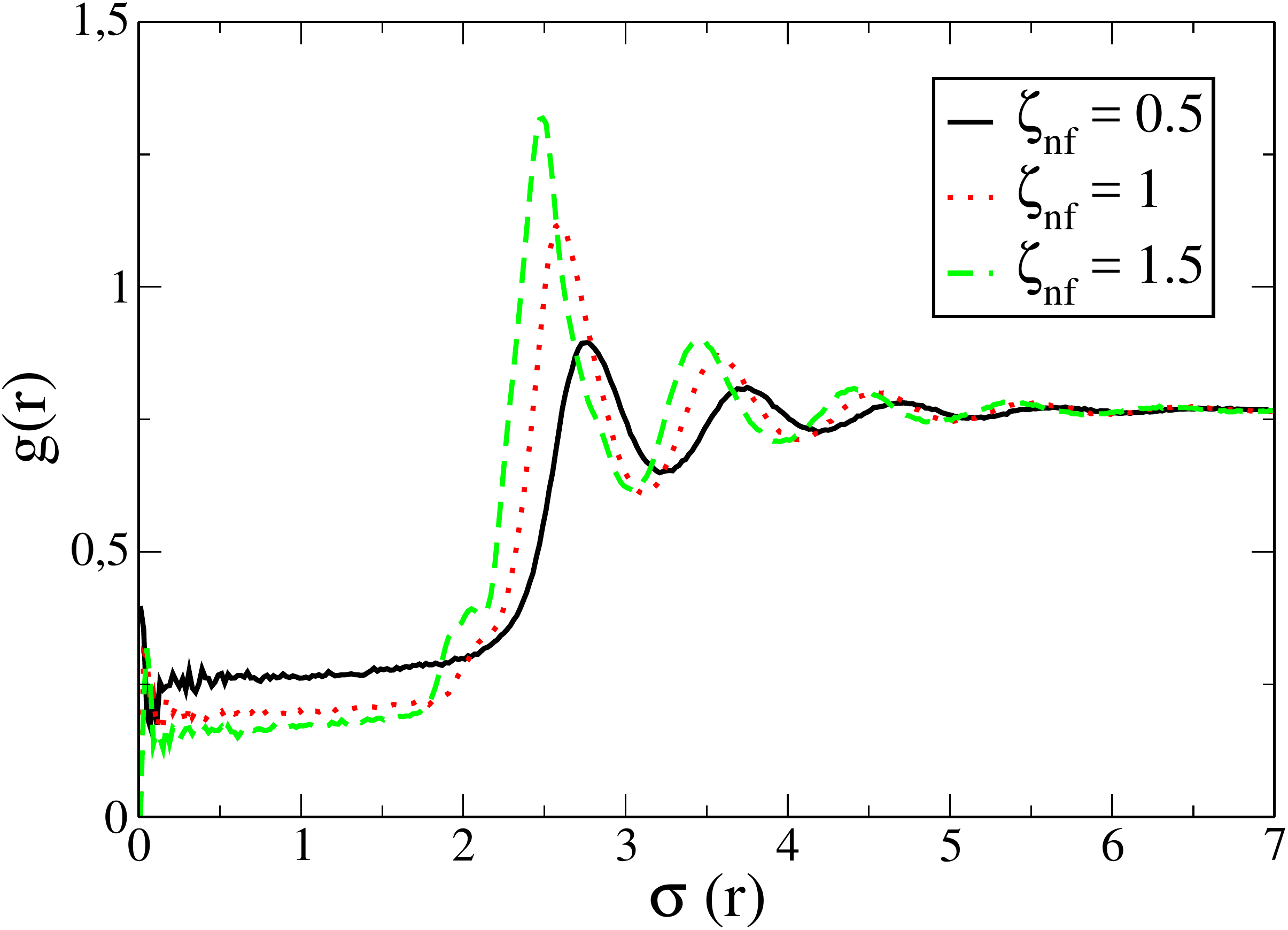}
 \caption{(Colour online) Radial distribution function of a nanoparticle for different values of the hydrophilic parameter $\zeta_{nf}$.}
 \label{rdf}
\end{figure}

We first investigate the effect of the hydrophilicity parameter $\zeta_{nf}$ on the density of the fluid in
the vicinity of a nanoparticle. We depict in Fig. \ref{rdf} the radial distribution function of a nanoparticle
for different values of the hydrophilic parameters $\zeta_{nf}$. The thickness of the layer is found to be approximately
$1 \sigma$ independently of $\zeta_{nf}$. However the liquid density in the layer is twice as large for $\zeta_{nf} = 1.5$ 
compared to $\zeta_{nf} = 0.5$. We observe that the hydrophilic interactions
between the particles and the fluid results in important layering effects for large values of $\zeta_{nf}$.

\begin{figure}[h!] 
        \includegraphics[width=9cm]{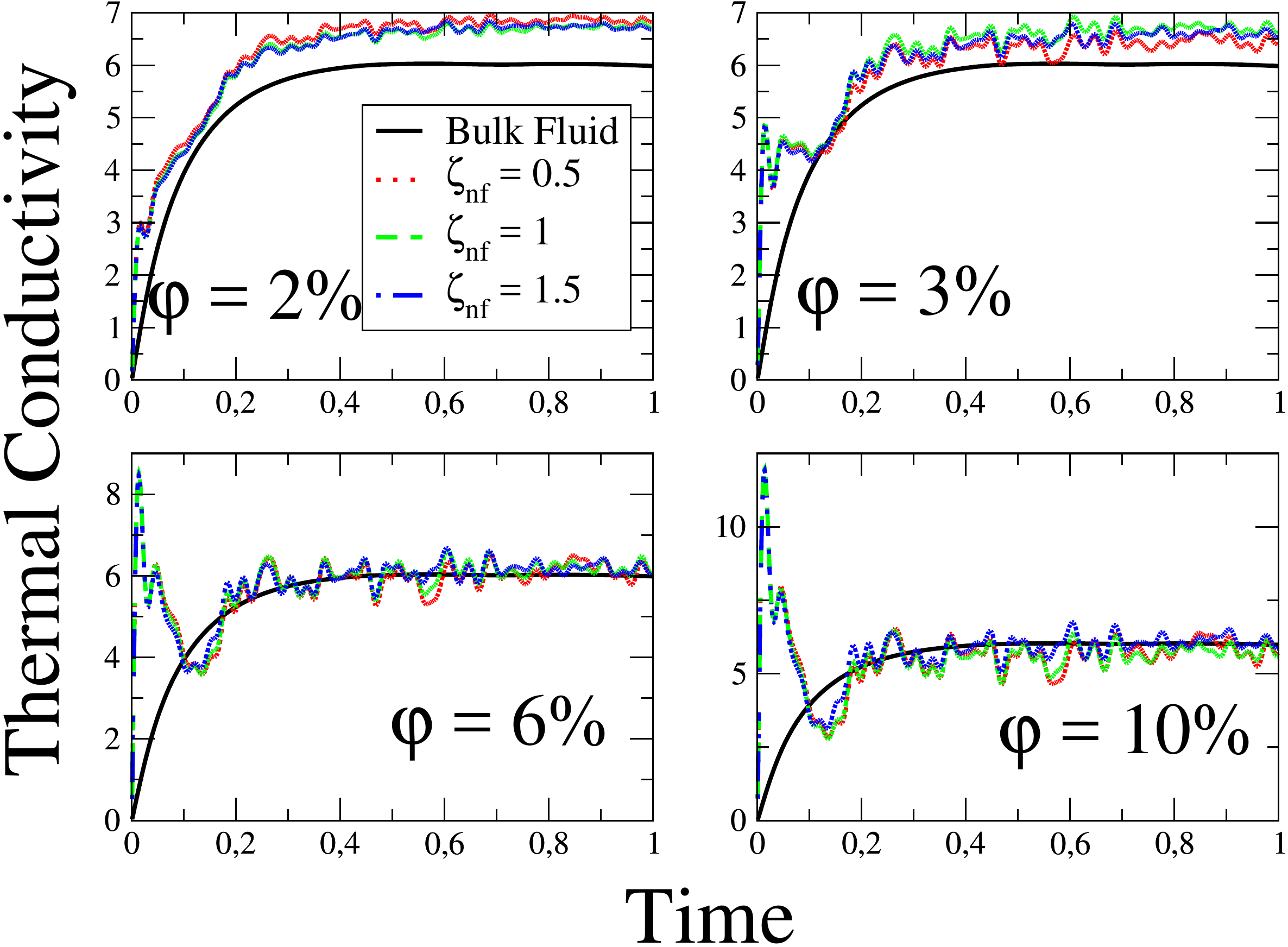}
        \caption{(Colour online) Integral of the heat current auto-correlation function for various nanofluids.}
        \label{nfl}
\end{figure}

We depict in Fig. \ref{nfl} the integral of the heat-current time auto-correlation function for different densities of
nanoparticles and varying hydrophilic parameter $\zeta_{nf}$.

In order to study the effect of the fluid
density distribution around the nanoparticle on the thermal conductivity of a nanofluid, the thermal
conductivity is evaluated for $\zeta_{nf} = 1.0$ and $\zeta_{nf} = 1.5$.

In the case of more than one nanoparticle, $\zeta_{nf} = 0.5$ is used to investigate the clustering effect on the
thermal conductivity of nanofluids.

We can rule out the layering effect since the hydrophilic parameter do not have an observable
effect on the thermal conductivity, as stated in the previous work by Xue \emph{et al.} \cite{30}.
Since the layering of the base fluid is directly related to the dispersion of the nanoparticles we can also
conclude that the aggregation of nanoparticles does not have an observable effect on the thermal conductivity in our model.
This result is in agreement with the study of Sedighi \emph{et al.} \cite{P3}
The results of all the nanofluid models demonstrate that there is no effect of particle-fluid interaction
on thermal conductivity, except a slight increase for small volume fractions.

We evaluate the thermal conductivity coefficients by computing a time average of the data on the
interval $0.4-1$. The thermal conductivity coefficient is found to be
approximately $6.7$ for a single nanoparticle and for $2$ nanoparticles in dimensionless units,
slightly higher than the bulk fluid. The thermal conductivity of nanofluids containing $4$ and $6$
nanoparticles are found as $6.1$ and $5.7$.

\begin{figure}[h!]
 \includegraphics[width=4.3cm]{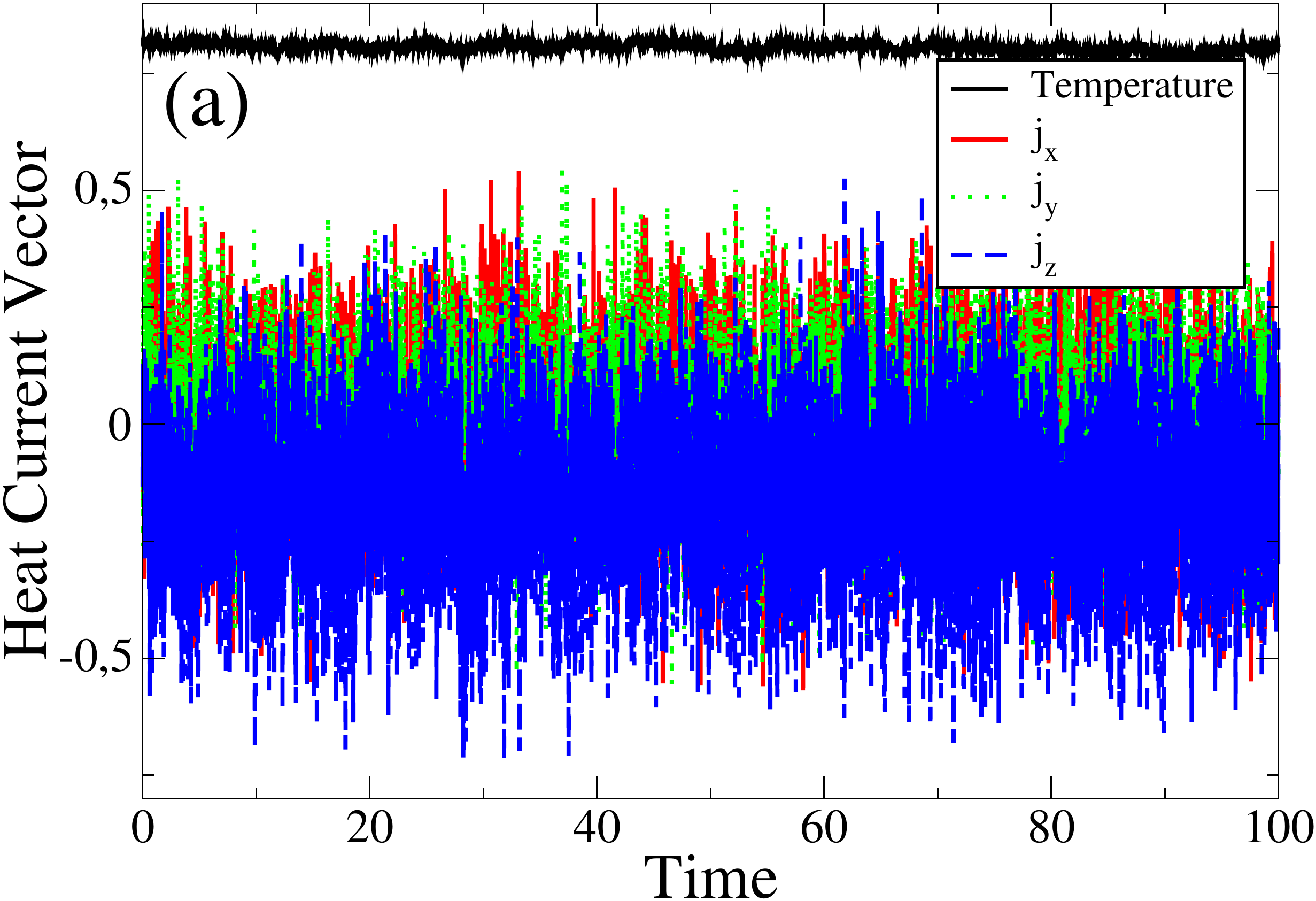}
 \includegraphics[width=4.1cm]{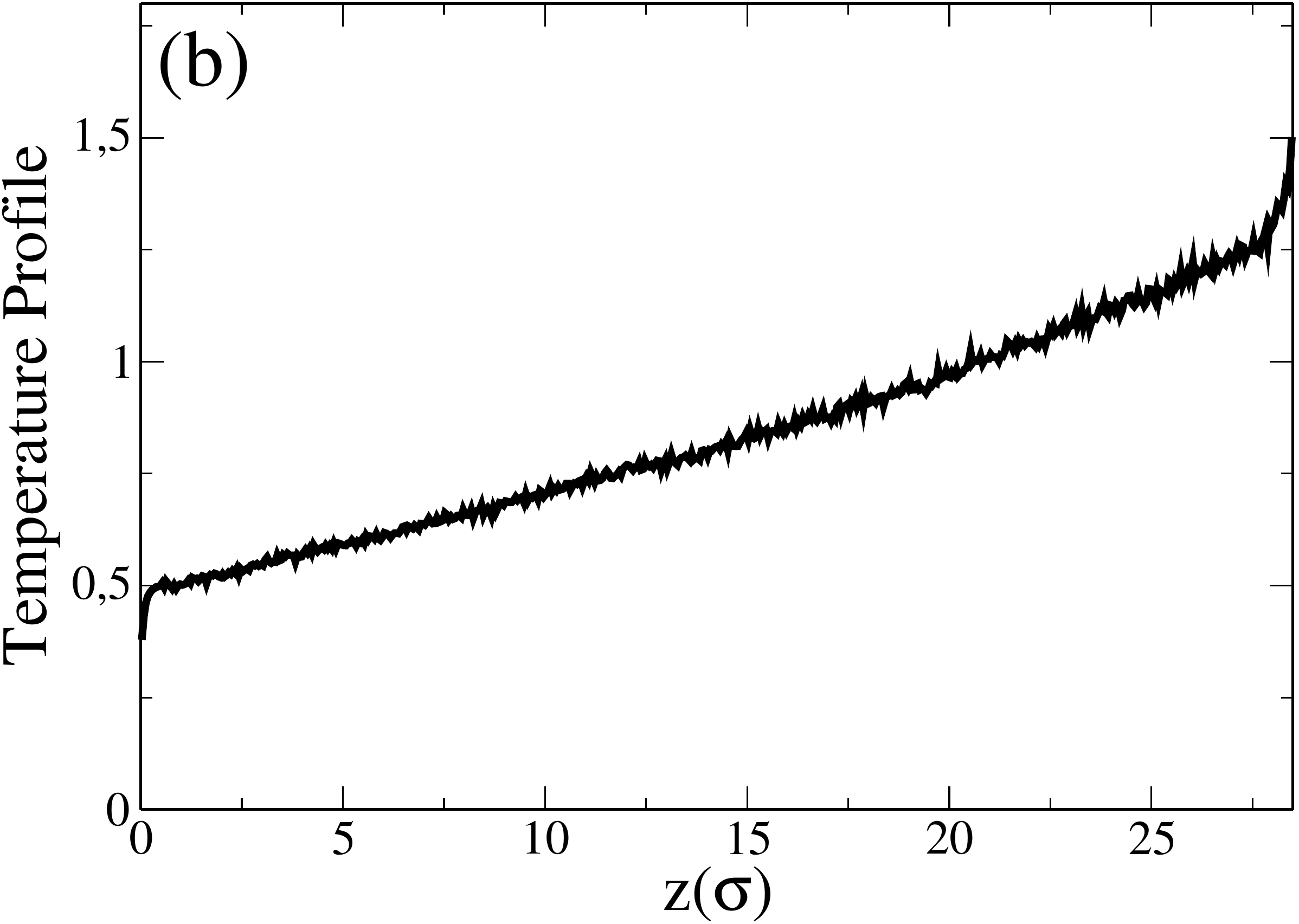}
 \caption{(Colour online) (a)Temperature and heat current for all three directions for the nanofluid consisting $1$ nanoparticle 
 during a NEMD simulation. (b)The temperature profile for $\Delta T=1.5$ as a function of $z$.}
 \label{nnemd}
\end{figure}

As in the case of the NEMD simulations, the heat current vector and the temperature profile for the nanofluid
with nanoparticle volume fraction $2\%$ are depicted in Fig. \ref{nnemd}, as an example.

The heat current auto-correlation functions of different nanofluids
has an oscillationary behavior. We observe that with increased volume fraction of nanoparticles oscillation
of the function is found to increase because of the long range correlations, as stated in the previous works of
Sarkar \emph{et al.} \cite{9} and Lee \emph{et al.} \cite{35}.

\begin{figure}[t!]
 \centering
 \includegraphics[scale=0.3]{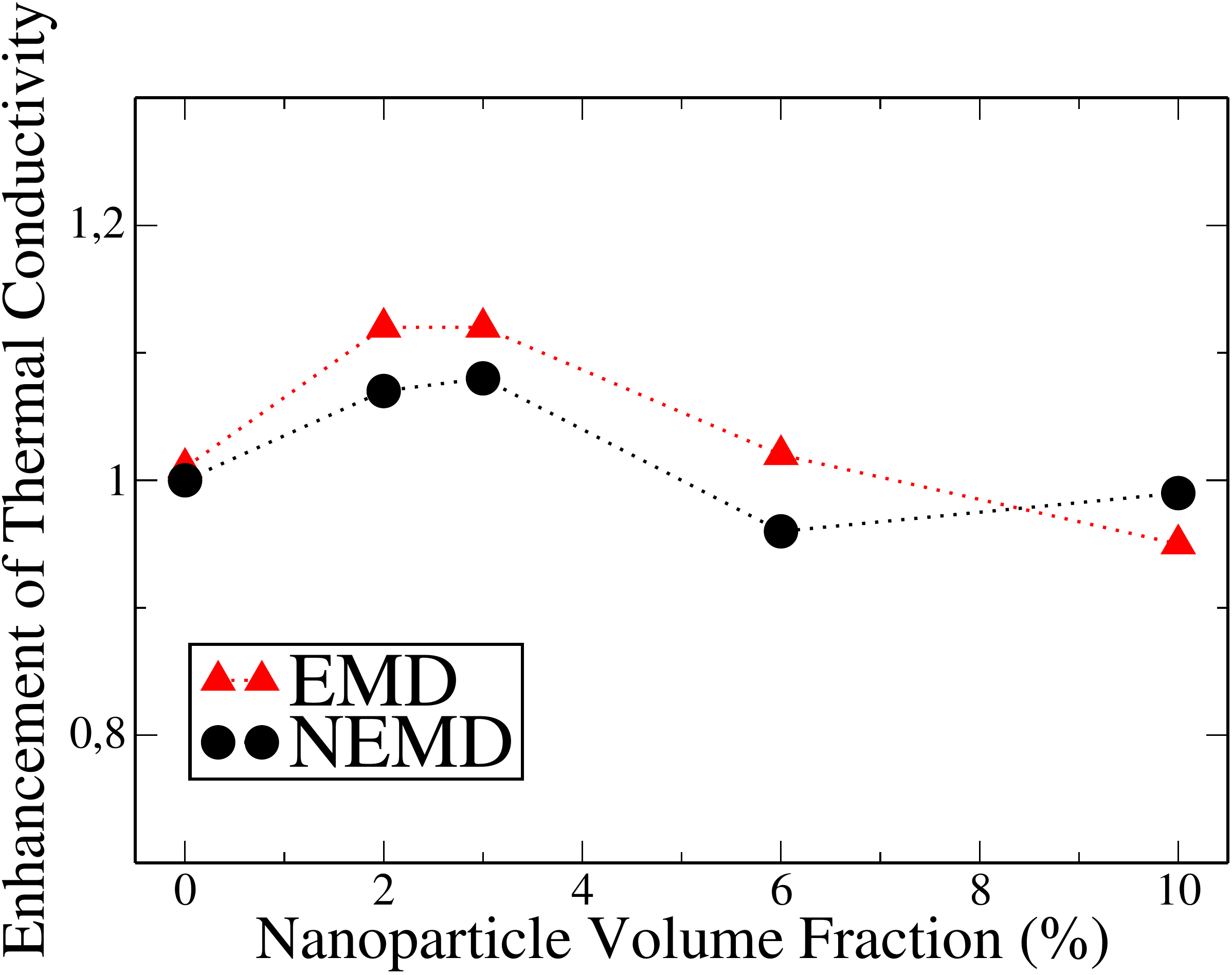}
 \caption{Thermal conductivity enhancement as a function of the volume fraction of nanoparticle.}
  \label{nevse}
\end{figure}

We depict the thermal conductivity enhancement as a function of the volume fraction obtained from both EMD and
NEMD calculations in Fig. \ref{nevse}. The results of NEMD calculations are in agreement with the EMD.
For low volume fractions ($2-3\%$) the thermal conductivity is found to increase of approximately $10\%$ with respect to the 
bulk fluid. For higher volume fractions, do not show any more enhancement, and surprisingly even decreases towards its bulk value.

\section{\label{sec5} Conclusions}

In this work we studied the thermal conductivity of nanoparticle and fluid composites. 
We modelled the base fluid with point particles interacting through Lennard-Jones interactions. 
The atoms inside the nanoparticles interact with a Lennard-Jones interaction, while the bonded pairs additionally interact with an attractive potential, the so called  FENE potential. 
The nanoparticle-nanoparticle, and nanoparticle-fluid interaction are also Lennard-Jones interaction with an additional term to control the strength of the attractive part. 
Both EMD and NEMD were carried out for systems with different volume fractions of nanoparticles.\par

The results show that the aggregation of nanoparticles does not affect the thermal conductivity significantly. 
Indeed, for three different values of the strength of the attractive part of the nanoparticle-fluid interaction we found the same thermal conductivity. 
This result is in agreement with previous works who ruled out the effect of liquid layering on thermal conductivity \cite{30}. On the other hand, 
the computations show that for low volume fractions of nanoparticles $(2-3\%)$ there is an increase of the thermal conductivity of a maximum value
of approximately  $10\%$ with respect to the bulk fluid. However, for larger volume fractions of nanoparticles there is no more enhancement, and even a decrease 
towards the bulk value of the conductivity at about $10\%$ volume fraction.\par

In order to check that our results are not size dependent we increased the number of fluid atoms and nanoparticles by fixing the volume fraction of nanoparticles 
and the fluid density.
The number of fluid atoms was increased from $5000$ to $15240$ and obtained two nanofluids with volume fractions $2\%$ and $6\%$ containing $4$ and $11$ nanoparticles respectively.
We observe the same results for thermal conductivity.\par

The lack of increase in the thermal conductivity of nanoparticle and fluid composites suggests that a number of previous numerical studies tend to over-estimate
the conductivity of the nanofluid by an incorrect definition of the average energies of the atoms. Indeed, we observed very large increases of the conductivity when the average
energies of the atoms of the nanoparticles were not carefully computed. We suggest that part of the discrepancies found in the literature could be due to this problem. \par

In order to confirm our results and find a reasonable explanation for the halt in the increase of the conductivity at higher volume fractions more extensive simulations are required. 
The size and shape of nanoparticles should be studied to rule out any system specific case and finite size effects. 
Electrostatic interactions should be taken into account to quantify the importance of long range interactions.

\end{document}